# Simulation of pulsed dynamic nuclear polarization in the steady state

Shebha Anandhi Jegadeesan[1], Yujie Zhao[2,3], Graham M. Smith[2],
Ilya Kuprov[4,5], and Guinevere Mathies[1,*]

[1]Department of Chemistry, University of Konstanz, Konstanz, Germany

[2]School of Physics and Astronomy, University of St Andrews, St Andrews, United Kingdom

[3]Current address: Francis Bitter Magnet Laboratory and Department of Chemistry,
Massachusetts Institute of Technology, Cambridge, Massachusetts, United States

[4]Chemical and Biological Physics Department, Weizmann Institute of Science, Rehovot, Israel

[5]School of Chemistry and Chemical Engineering, University of Southampton,
Southampton, United Kingdom

## Abstract

In pulsed dynamic nuclear polarization (DNP), enhancement of the polarization of bulk nuclei requires the repeated application of a microwave pulse sequence. So far, analysis of a one-time transfer of electron spin polarization to a dipolar-coupled nuclear spin has guided the design of DNP pulse sequences. This has obvious shortcomings, such as an inability to predict the optimal repetition time. In an actual pulsed DNP experiment, a balance is reached between the polarization arriving from the unpaired electrons and nuclear relaxation. In this article, we explore three algorithms to compute this stroboscopic steady state: (1) explicit time evolution by propagator squaring, (2) generation of an effective propagator using the matrix logarithm, and (3) direct calculation of the steady state with the Newton-Raphson method. Algorithm (2) is numerically unstable for this purpose. Algorithm (1) and (3) are both stable; algorithm (3) is the most efficient. We compare the steady-state simulations to existing experimental results at 0.34 T and 1.2 T and to the first experimental observation of X-inverse-X (XiX) DNP at 3.4 T: agreement is good, and improves further when electron-proton distance and electron Rabi frequency distributions are accounted for. We demonstrate that the trajectory of the spin system during one-time application of a microwave pulse sequence differs from the steady orbit. This has implications for DNP pulse sequence design.

*Email: guinevere.mathies@uni-konstanz.de

# 1. Introduction

A well-chosen microwave pulse sequence readily transfers electron spin polarization to a nearby nuclear spin. In pulsed dynamic nuclear polarization (DNP), the repeated application of such a microwave pulse sequence generates enhanced polarization of bulk nuclei (Figure 1a). This approach may be a valuable alternative to classical continuous-wave DNP,[1–3] because it can provide flexible excitation of the electron spins as well as a more efficient polarization transfer. Of particular interest is the use of pulsed DNP to enhance the sensitivity of high-resolution magic-angle spinning nuclear magnetic resonance (MAS NMR),[4,5] but to make this possible several technical challenges have to be addressed. Most significantly, DNP pulse sequences require a high electron Rabi frequency ($\omega_{1S} = -\gamma_e B_1$, with $\gamma_e$ the electron gyromagnetic ratio and $B_1$ the strength of the magnetic field component of the microwave irradiation), on the order of $2\pi \cdot 10$ MHz at least. Depending on MAS rotor size and conversion factor of the probe, this means an estimated peak microwave power of 100 W or more. Above 95 GHz (the electron Larmor frequency at 3.4 T; the corresponding proton Larmor frequency is 144 MHz), coherent sources with such output power are challenging to make.

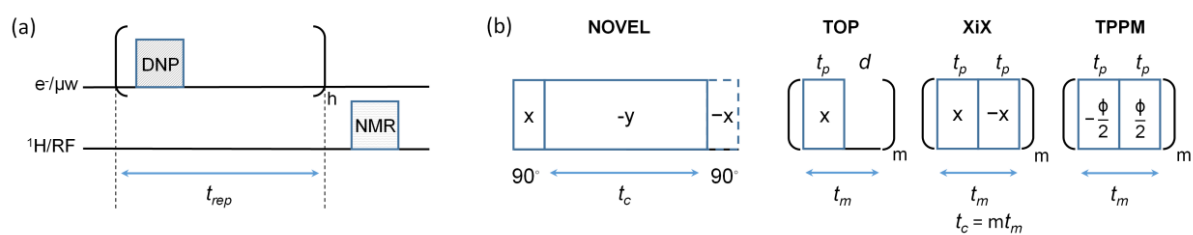

*Figure 1.* (a) General scheme of pulsed DNP. One instance of a DNP pulse sequence transfers electron spin polarization from a paramagnetic species or polarizing agent to a nearby nucleus, typically a proton. To build up enhanced polarization of bulk nuclei, the sequence is applied hundreds to thousands of times. The optimal repetition time, $t_{rep}$, typically on the order of a millisecond, depends on the pulse sequence and relaxation properties of the chemical system. (b) DNP pulse sequences consisting of monochromatic, constant amplitude pulses. Polarization transfer occurs at the matching condition. For Nuclear spin Orientation Via Electron spin Locking (NOVEL), the electron Rabi frequency (also referred to as the nutation frequency), $\omega_{1S}$, has to match the nuclear Larmor frequency, $\omega_{0I}$. The possibility of microwave irradiation off resonance ($\Omega \neq 0$), is accommodated in the general matching condition $|\omega_{eff}| = \sqrt{\omega_{1S}^2 + \Omega^2} = |\omega_{0I}|$. For Time-Optimized Pulsed (TOP), X-inverse-X (XiX), and Two-Pulse Phase Modulation (TPPM) DNP, a multiple of the modulation frequency, $\omega_m = 2\pi/t_m$, plus or minus the effective frequency, $\omega_{eff} = \beta/t_m$, with $\beta$ the effective flip angle of the electron spin magnetization due to each block of the sequence, has to match the nuclear Larmor frequency: $k\omega_m + l\omega_{eff} + n\omega_{0I} = 0$, with $k = \pm 1, \pm 2, ...$ and $l, n = \pm 1$. The contact time, $t_c$, varies from a few hundred nanoseconds to a few microseconds, depending on the sequence.

As one approaches the millimeter wave regime, solid-state sources and amplifiers are no longer able to produce high-power irradiation.[6] For wavelengths up to about a millimeter, slow-wave vacuum devices such as the klystron are a viable alternative[7] and are currently successfully used in high-resolution MAS DNP at 9.4 T/400 MHz/263 GHz.[8] At still higher fields and frequencies, interaction structures (of the dimension of the wavelength) become too fine and incompatible with the generation of high-power irradiation. Instead, overmoded, fast-wave vacuum devices must be used. For this reason, dedicated gyrotrons were developed as the sources of continuous high-frequency, high-power (typically tens of watts) irradiation in MAS DNP up to 21.1 T/900 MHz/592 GHz.[9,10] Generating nanosecond or picosecond *pulses* with fast-wave devices, however, comes with its own difficulties. Phase and frequency stability, for example, are notoriously difficult to maintain. A strategy to overcome this is to first generate coherent pulses at low power (milliwatts to watts) using a solid-state source and amplify them (by about 30 dB) using a fast-wave device.[11] Prototypes that follow this strategy have been built.[12,13] Direct generation at high power is also being explored.[14]

Recent years have seen a surge in the development of DNP pulse sequences and their experimental implementations at low magnetic field. They can be broadly classified into: (1) pulse sequences with monochromatic pulses of fixed amplitude and phase that meet a matching condition (see Figure 1b for an overview) and (2) pulse sequences that rely on adiabatic passage through a matching condition and therefore typically include a frequency or amplitude sweep.[15–17] NOVEL[18] and (without the initial 90° pulse) pulsed solid-effect[19] were the first sequences in the former category. The TOP/XiX/TPPM DNP and also the Broadband Excitation by Amplitude Modulation (BEAM) sequence[20–23] comprise a next generation and have the advantage that the matching condition can be met even if the electron Rabi frequency is well below the nuclear Larmor frequency. However, the enhancement and build-up time of the nuclear polarization still depend on the sequence in combination with the available peak microwave power.

We recently investigated the TOP, XiX, and TPPM DNP sequences both theoretically and experimentally at 1.2 T/34 GHz/51 MHz.[22] To enable a fair evaluation of performance, the parameters for each sequence had to be carefully optimized. We started by numerically scanning the quality of possible DNP conditions across all experimentally feasible microwave resonance offsets, electron Rabi frequencies, pulse lengths, and (for TPPM) phases. To this end, we simulated a single (one-time) transfer of polarization from an electron to two nearby protons, brought about by one instance of the DNP pulse sequence. The presence of the second proton was not essential, but improved the agreement with experimental line shapes. The best conditions were subsequently tested on the spectrometer. Predicted parameters were adjusted if necessary and, in addition, the contact and repetition times were optimized. Following this procedure, we found that TPPM DNP, at an electron Rabi frequency of 33 MHz, generated both the highest enhancement factor of the proton polarization and the fastest build-up. XiX DNP was the runner-up, but worked very well at a much lower electron Rabi frequency of 7 MHz. Generally, we found that a better preservation of the electron-nuclear dipolar coupling in the effective Hamiltonian of the DNP sequence produces faster build-up of the bulk dynamic nuclear polarization, but does not imply a higher enhancement factor.[22]

A further conclusion from the investigations at 1.2 T was that simulations of the single transfer could not reproduce all experimental observations: peak shapes and intensities in field profiles differed between simulations and experiments, as did contact curves and the dependence on the electron Rabi frequency. Effects of the repetition time could obviously not be simulated. Thus, better simulation tools are needed for the evaluation of DNP pulse sequences and to guide the design of microwave sources.

Two numerical complexity obstacles[24] arise in the simulation of the full (pulsed) DNP process. The first is that very large spin systems have to be considered. Each polarizing agent serves hundreds to thousands of nuclear spins. To capture the build-up of the bulk nuclear polarization, all these spins and their interactions have to be modelled. In addition, we know that interactions between polarizing agents matter.[22] A first, critical step towards solving this problem was taken with the introduction of restricted state spaces.[24,25] For the large majority of magnetic resonance contexts (the single crystal is the notable exception)[26], it suffices to consider 4-5 spins per product state.[27] This approach has been used to simulate the static, continuous-wave solid-effect[28] as well as the cross-effect under MAS[29,30]. Nevertheless, to model nuclear spin diffusion, additional measures remained necessary: in the static solid-effect, the build-up of bulk polarization has been modelled as kinetically constrained diffusion,[31] in cross-effect under MAS, the Landau-Zener approximation is used to determine the polarization transfer during rotor events.[32,33] More recently, the realization that state spaces can be further restricted by using a neighbor's cutoff[34] has enabled large-scale *ab initio* simulation of MAS DNP.[35]

The second obstacle is that the steady state (see below for the formal definition) is reached only when electron spin polarization, nuclear spin diffusion, and relaxation are balanced; this can take seconds. In MAS DNP, this is after many rotor periods; in pulsed DNP, this is after the DNP sequence has been applied many times. In a brute force simulation approach, the spin system is propagated explicitly through these cycles. This is computationally expensive: in their simulations of MAS DNP, Perras and Pruski had to use a Monte Carlo method to avoid storing the propagator.[29] In this article, we set out to efficiently calculate the steady state.

Periodically-driven, dissipative linear systems do not, in general, have a steady state. Instead, they have a steady orbit – a periodic trajectory to which system dynamics eventually converge.[36] A stroboscopic steady state may then be defined as a state at a given point of the steady orbit, usually the point that corresponds to the end of the period. We explore three algorithms to calculate stroboscopic steady states numerically from the propagator of the complete pulsed-DNP period, i.e., the repetition time including the pulse sequence (Figure 1a). The first algorithm uses repeated squaring of the propagator matrix. Every squaring operation doubles the time step; a high enough power of the propagator represents long-term evolution well. This always works, but is numerically expensive. The second algorithm computes an effective evolution generator using the matrix logarithm method,[37] puts the time derivative to zero in the Liouville-von Neumann equation, and solves the resulting algebraic equation for the steady state. This method works well in NMR spectroscopy[38], but large and badly conditioned matrices make it numerically unstable in dissipative DNP settings. The third algorithm, which we advocate here, uses the Newton-Raphson method to find the stroboscopic steady state from the pulsed-DNP propagator in just a few matrix-vector operations.

We demonstrate steady-state simulations of NOVEL at 0.34 T and of TOP, XiX, and TPPM DNP at 1.2 T. We also present the first experimental results of XiX DNP at 3.4 T, along with corresponding steady-state simulations. Experimental field profiles, optimization of the repetition times, and the dependence on the electron Rabi frequency are remarkably well reproduced, even when just one electron and one proton are considered. The same is true for the polarization transfer during the contact time, which provides evidence that steady orbits differ from the trajectories of the spin system during the first polarization transfer. Integration over the proton position distribution (including its effect on the nuclear relaxation) and the distribution in the microwave $B_1$-field improves the agreement with experiments further.

## 2. Stroboscopic steady-state algorithms

A standard result from the dynamical systems theory[36] is that a periodically driven dissipative linear system (a) always has a unique steady orbit and (b) converges to that orbit from any initial condition. Spin system dynamics under a DNP pulse sequence is dissipative, linear, and periodically driven – this is visible in the equation of motion for the density matrix $\boldsymbol{\rho}$:

$$\frac{\partial \boldsymbol{\rho}}{\partial t} = -i\left(\mathcal{H}_0 + \mathcal{H}_1(t)\right)\boldsymbol{\rho} + \mathcal{R}\left(\boldsymbol{\rho} - \boldsymbol{\rho}_{eq}\right) \tag{1}$$

which has the time-independent interactions in the drift Hamiltonian commutation superoperator $\mathcal{H}_0$, the time-periodic pulse sequence events in $\mathcal{H}_1(t)$, and the negative-definite relaxation superoperator $\mathcal{R}$ that drives the system towards the thermal equilibrium state $\boldsymbol{\rho}_{eq}$. The latter part is sometimes abbreviated to use a "thermalized" relaxation superoperator $\mathcal{R}_\theta \boldsymbol{\rho} = \mathcal{R}\left(\boldsymbol{\rho} - \boldsymbol{\rho}_{eq}\right)$. A unique steady orbit therefore exists; a *stroboscopic* steady state $\boldsymbol{\rho}_\infty$ may be defined as the system state at the point in the steady orbit that matches the pulsed-DNP period, i.e., the repetition time including the pulse sequence (Figure 1a). Once $\boldsymbol{\rho}_\infty$ is computed, the steady orbit may be obtained by propagating $\boldsymbol{\rho}_\infty$ through the period.

## 2.1 Propagator squaring

Consider the propagator $\mathcal{P}$ taking the system forward in time by one pulsed-DNP period $T$:

$$\begin{aligned}\boldsymbol{\rho}(t+T) &= \mathcal{P}\boldsymbol{\rho}(t) \\ \mathcal{P} &= \lim_{\Delta t_k \to 0} \overset{\leftarrow}{\prod_k} \exp\left\{\left[-i\left(\mathcal{H}_0 + \mathcal{H}_1(t_k)\right) + \mathcal{R}_\theta\right]\Delta t_k\right\}\end{aligned} \tag{2}$$

Despite its formidable-looking definition (the limit of time-ordered products of thin time slice propagators) $\mathcal{P}$ is easy to compute,[39,40] particularly when the Hamiltonian is piecewise-constant in the interaction representation.[41] The properties discussed above guarantee that repeated action by $\mathcal{P}$ on any physically valid (meaning Hermitian, non-negative definite, unit trace) density matrix $\boldsymbol{\rho}_0$ would eventually yield the stroboscopic steady state:

$$\boldsymbol{\rho}_\infty = \lim_{n \to \infty}\left(\mathcal{P}^n \boldsymbol{\rho}_0\right) = \mathcal{P}\left(\mathcal{P}\left(\cdots\left(\mathcal{P}\left(\mathcal{P}\boldsymbol{\rho}_0\right)\right)\right)\right) \tag{3}$$

However, a practical difficulty is that the pulsed-DNP period contained in $\mathcal{P}$ is in the milliseconds, whereas the nuclear relaxation times in $\mathcal{R}$ that determine the steady orbit settling time are in the seconds. To reach the steady state, $\mathcal{P}$ therefore has to be applied thousands of times. This is computationally expensive, even if $\boldsymbol{\rho}_0$ is carefully chosen.

One simple solution is to observe that every time a propagator is squared, its evolution time doubles[39,40], the settling time may therefore be reached by performing around $\log_2 10^3 \approx 10$ sequential squaring operations; this is particularly efficient on a graphics processing unit (GPU):

$$\boldsymbol{\rho}_\infty = \lim_{n \to \infty}\left(\mathcal{P}^{2^n}\boldsymbol{\rho}_0\right) = \left(\left(\left(\mathcal{P}^2\right)^2\right)^{\cdots}\right)^2 \boldsymbol{\rho}_0 \tag{4}$$

In practice, propagator squaring is performed until the next squaring no longer changes the resulting state to a user-specified tolerance; numerical efficiency trade-offs in such procedures are explored in our recent work.[42] This method is implemented as the 'squaring' option in the function steady.m of *Spinach* 2.10 and later versions.[43] In magnetic resonance settings, it is unconditionally stable. However, it still involves a significant number of matrix-matrix multiplications in Liouville space – this is better than the naïve time evolution method, but still expensive. A small blessing is that sequential squares of the propagator do eventually become sparser; sparse index clean-up[41] after each square is therefore recommended.

## 2.2 Effective Hamiltonian

An alternative method for computing the stroboscopic steady state is to find an effective Hamiltonian that describes the net result of the events inside the pulsed-DNP period. Consider the equation of motion expressed via the effective Hamiltonian $\bar{\mathcal{H}}$ over the pulsed-DNP period $T$:

$$\frac{\partial \boldsymbol{\rho}}{\partial t} = -i\bar{\mathcal{H}}\boldsymbol{\rho} + \bar{\mathcal{R}}\left(\boldsymbol{\rho} - \boldsymbol{\rho}_{eq}\right) \qquad (5)$$

Such equations, and ways of computing $\bar{\mathcal{H}}$ are extensively researched,[44,45] their solutions are stroboscopically correct – they coincide with the solution of Eq. (1) at the edges of the pulsed-DNP period. Eq. (5) may therefore be used to obtain the stroboscopic steady state: not the entire orbit, but the state at the end of each orbital period. From the definition of the steady state (zero-time derivative in the equation of motion), we obtain a straightforwardly computable expression:

$$\mathbf{0} = -i\bar{\mathcal{H}}\boldsymbol{\rho}_\infty + \bar{\mathcal{R}}\left(\boldsymbol{\rho}_\infty - \boldsymbol{\rho}_{eq}\right) \quad \Rightarrow \quad \boldsymbol{\rho}_\infty = \left(-i\bar{\mathcal{H}} + \bar{\mathcal{R}}\right)^{-1}\bar{\mathcal{R}}\boldsymbol{\rho}_{eq} \qquad (6)$$

The lower-upper (LU) preconditioned generalized minimum residuals method (GMRES)[46] is recommended for the inverse-times-vector operation – the matrix inverse in Eq. (6) need not be calculated explicitly.

In practical calculations, the effective Hamiltonian method has three logistical problems, which we call the good, the bad, and the ugly. The good problem is the calculation of the effective evolution generator – analytical series converge slowly[47] and the numerical method for computing the propagator logarithm[37]

$$\bar{\mathcal{H}} + i\bar{\mathcal{R}} = \frac{i}{T}\ln\left(\mathcal{P}\right) \qquad (7)$$

requires repeated matrix square roots that rely on factorizations that have no efficient sparse, parallel, or GPU-based implementations. For small spin systems in Hilbert space, the matrix logarithm method works beautifully[48] because all evolution generators are Hermitian (meaning diagonalization is always possible), all propagators are unitary (meaning the best possible condition number of 1), and interaction representations keep evolution generator condition numbers benign. However, $\bar{\mathcal{H}} + i\bar{\mathcal{R}}$ may be ill-conditioned; this is the bad problem. Even in the rotating frame, the largest eigenvalues of $\bar{\mathcal{H}}$ are in the tens of MHz (hyperfine couplings and electron nutation frequencies) and the smallest ones are actually zero because $\bar{\mathcal{H}}$ is a commutation superoperator. One must then rely on the smallest absolute eigenvalue of the negative definite matrix $\bar{\mathcal{R}}$ for stability, which may be less than 0.1 Hz (bulk nuclei in a cryogenic sample). This then yields a condition number (the ratio of the largest eigenvalue to the smallest one) for $\bar{\mathcal{H}} + i\bar{\mathcal{R}}$ of $10^8$ that gets worse (because multiple-quantum frequencies appear) with each additional particle. The result is that the inverse-times-vector operation in Eq. (6) will not fit into the double floating point format (precision of $2^{-53} = 1.11 \times 10^{-16}$), making it numerically unstable. Finally, the ugly problem is that $\bar{\mathcal{H}}$ and $\bar{\mathcal{R}}$ do not, in general, commute. This then requires a formalism upgrade from the group-theoretically sound average Hamiltonian theory to its treacherous semigroup equivalent describing dissipative dynamics.

## 2.3 Newton-Raphson

By its definition, the stroboscopic steady state does not change under the action by the pulsed-DNP propagator $\mathcal{P}$:

$$\mathscr{P}\boldsymbol{\rho}_\infty = \boldsymbol{\rho}_\infty \quad \Rightarrow \quad \begin{cases} (\mathscr{P}-\mathbf{1})\boldsymbol{\rho}_\infty = \mathbf{0} \\ \text{Tr}(\boldsymbol{\rho}_\infty) = 1 \end{cases} \qquad (8)$$

This is a fixed-point finding problem with a linear constraint; it may be solved using the Newton-Raphson root-finding algorithm.[49] The spherical tensor basis[50] used by *Spinach* in Liouville space is particularly convenient here because the $\mathbf{T}_{00}$ coefficient corresponding to the density matrix trace is the first element $\rho_1$ of the state vector[50] and in thermodynamically consistent simulations this element is always equal to 1. We therefore only need to solve Eq. (8) for the rest of the state vector. The procedure is as follows:

1. Set up the residual vector that is to be driven to zero: $\mathbf{f}(\boldsymbol{\rho}) = (\mathscr{P}\boldsymbol{\rho} - \boldsymbol{\rho})_{2:n}$.

2. Form the Jacobian $\mathbf{J} = \mathscr{P}_{2:n,2:n} - \mathbf{1}$. Optionally, pre-compute its LU factorization to facilitate the subsequent repeated inverse-times-vector operations.

3. Repeatedly take the Newton-Raphson step $\boldsymbol{\rho}^{(k+1)}_{2:n} = \boldsymbol{\rho}^{(k)}_{2:n} - \mathbf{J}^{-1}\mathbf{f}(\boldsymbol{\rho}^{(k)})$ until there is no change in the state vector to the user-specified tolerance ($10^{-8}$ or smaller). LU-preconditioned GMRES[46] method is recommended here.

A software implementation is available under the 'newton' option in the function steady.m of *Spinach* 2.10 and later versions.[43] The LU factorization may be pre-computed because the Jacobian is static.

The Newton-Raphson procedure does not suffer from numerical stability problems, because all eigenvalues of the propagator are inside the unit circle but well away from zero. The eigenvalues of the Jacobian are therefore all of the form $e^{-i\omega T}e^{-rT} - 1$ where $\omega$ and $r$ are positive real numbers, $T$ is the pulsed-DNP period, and $rT \ll 1$. Upper and lower bounds on the modulus of this quantity are:

$$r_{\min}T \leq |e^{-i\omega T}e^{-rT} - 1| \leq 2 \qquad (9)$$

where $r_{\min}$ is the smallest relaxation rate found in the system, on the order of 0.1 Hz for longitudinal relaxation of bulk nuclei. A typical pulsed-DNP period is in the milliseconds, yielding a condition number $2/r_{\min}T$ of around $10^4$ for the Jacobian; this is easily handled in double precision arithmetic. Importantly, this condition number does not depend on the system size.

## 3. Materials and methods

### 3.1 Experimental

All pulsed DNP experiments were performed on samples of 6 mM of the narrow-line radical trityl OX063 (obtained from Polarize ApS) doped into a glassy matrix of $d_8$-glycerol:$D_2O$:$H_2O$ 60:30:10 v:v:v ("DNP juice") at a temperature of 80 K.[51] Data at 0.34 T/15 MHz/9.7 GHz (X band) and 1.2 T/34 GHz/51 MHz (Q band) are reproduced from Jain *et al.*[19] and Redrouthu *et al.*[22], respectively. At both fields, a commercial pulsed electron paramagnetic resonance (EPR) spectrometer was used, which had been extended in-house with an NMR console. The ENDOR (electron nuclear double resonance) probe was made suitable for $^1$H NMR excitation and detection with a custom-built tuning/matching box. At X band, microwave pulses were amplified to 1 kW using a travelling-wave tube (TWT) and used to excite a sample volume of 20-50 μL in a fully over-coupled dielectric resonator. At Q band, microwave pulses were amplified to 50 W using a solid-state amplifier and used to excite a sample volume of 1-2 μL, also in a fully over-coupled dielectric resonator. $^1$H NMR signals had to be detected with an echo sequence, because ring-down temporarily saturates the receiver after the radiofrequency pulse. A typical delay between the pulses of the echo sequence was 20-40 μs.

Data at 3.4 T/143 MHz/94 GHz (W band) were newly acquired on a pulsed EPR spectrometer constructed by Cruickshank *et al.*,[52] which was recently upgraded.[53] A pulsed extended-interaction klystron amplifier (EIKA) provides up to 1.3 kW of peak power across an operational range of 93.5-94.5 GHz. The non-resonant sample-holder supports a single transverse waveguide mode (active volume 30 μL) with a mean conversion efficiency of 1.4 MHz/W$^{1/2}$. For the pulsed DNP experiments reported here, the polarization of the microwave irradiation was converted from linear to circular, at a cost of reduced power, but in favor of a more homogeneous excitation across the sample (electron Rabi frequency of 20 MHz). $^1$H NMR signals were excited and detected using a locally-tuned saddle coil. After microwave irradiation, a radiofrequency π/2-pulse was applied and the free induction decay could be detected after a delay of 7 μs. For DNP-enhanced NMR signals, a single acquisition produces sufficient signal-to-noise ratio.

### 3.2 Hardware and software

Simulations were performed using *Spinach*[43] on a Dell PowerEdge T550 server equipped with two Intel Xeon 6326 Gold processors (32 cores in total), 512 GB of RAM, and an Nvidia A100 GPU (80 GB of RAM, 9.7 TFLOPS FP64). The initial state was thermal equilibrium. The transformation of longitudinal electron spin magnetization ($\mathbf{S}_Z$) into longitudinal nuclear spin magnetization ($\mathbf{I}_Z$) was monitored by reading out the expectation value according to

$$\langle \mathbf{I}_Z \rangle_\infty = \mathrm{Tr}(\mathbf{I}_Z \boldsymbol{\rho}_\infty) \tag{10}$$

Note that the longitudinal magnetization is half of the normalized population difference or polarization. For the sudden DNP sequences investigated in this work, we expect, after powder averaging, a maximum enhancement factor of $\frac{1}{2}|\gamma_e/\gamma_{^1H}|$ = 329. The corresponding values of $\langle \mathbf{I}_Z \rangle$ are 7.1×10$^{-4}$ at X band, 2.5×10$^{-3}$ at Q band, and 7.1×10$^{-3}$ at W band.

### 3.3 Model spin systems

The basic spin system for the steady-state simulations consists of one electron and one proton separated by 3.5 Å, corresponding to a dipolar coupling of 1.8 MHz. To reproduce the EPR line shape of the trityl OX063 radical, *g*-anisotropy was included ($g_x = g_y$ = 2.00319, $g_z$ = 2.00258). The electron-proton coupling was adjusted to reproduce the polarization transfer times observed in NOVEL and pulsed SE experiments at 0.34 T[19] and is clearly stronger than individual electron-proton couplings.[54,55] The reason is that the unpaired electron of trityl OX063 is surrounded by tens of protons at similar distances, which are simultaneously available for polarization transfer.[55–57] We briefly explored the placement of further protons in the proximity of the electron, but steady-state simulations of these larger spin systems were not better than those using ensemble averaging over the position of a single proton as described below. Still, for the sake of completeness, we did use chains of up to nine protons, starting at 3.5 Å with each subsequent proton 1.5 Å further away, for computational efficiency benchmarking of the steady-orbit algorithms. The dimension of the Liouville space was reduced by including only product states between up to a specified (see below) number of spins.[24,27]

### 3.4 Relaxation

Experimental information at Q band[22] was used as a guide for choosing spin relaxation times. At X band,[51] W band,[58,59] and 140 GHz[60], as far as we know, there are no major differences – for the purposes of this work, it was sufficient to set the orders of magnitude correctly. The longitudinal electron spin relaxation time $T_{1e}$ was measured, by inversion recovery, to be 1.8 ms; we set the value to 1 ms. In an echo decay experiment, the phase memory time was found to be 1.8 μs; since processes other than transverse relaxation contribute significantly to dephasing, we set $T_{2e}$ to 5 μs. The longitudinal relaxation time of the bulk protons $T_{1n,bulk}$ was found to be 52 s; we used this value at Q and W band. At X band, we took 26 s from Mathies *et al.*[51] Finally, we used the $^1$H spin echo decay time as an estimate for the transverse nuclear relaxation time, $T_{2n}$; we set it to 20 μs throughout, assuming that flip-flops with neighboring protons are the dominant mechanism, even for the protons nearest to the trityl radical.[61]

Close to the radical, longitudinal relaxation of nuclear spins becomes faster. Consider an isolated, non-mobile unpaired electron and its dipolar coupling to a nearby spin-1/2 nucleus. Longitudinal relaxation of the electron spin gives rise to fluctuations of the dipolar coupling with correlation time $T_{1e}$. Via the $C$ and $D$ terms of the dipolar alphabet,[62] these cause single-quantum nuclear spin transitions, resulting in longitudinal nuclear relaxation. When $\nu_{0I} T_{1e} \gg 1$,

$$\frac{1}{T_{1n}} \simeq \left(\frac{\nu_{dip}}{\nu_{0I}}\right)^2 \frac{1}{T_{1e}} \text{sech}^2\left(\frac{g_e \mu_B B_0}{2 k_B T}\right) \tag{11}$$

where $\nu_{dip}$ is the electron-nuclear dipolar coupling

$$\nu_{dip} = \frac{\mu_0}{4\pi} \frac{g_e g_n \mu_B \mu_N}{h} \frac{1 - 3\cos^2\theta}{r_{dip}^3} \tag{12}$$

and $\nu_{0I}$ is the nuclear Larmor frequency.[61] The hyperbolic secant accounts for the fact that when the electron spins are fully polarized, fluctuations disappear and this contribution to nuclear relaxation vanishes. All other symbols have their usual meaning: $g_e$ is the free-electron *g*-factor, $\mu_B$ is the Bohr magneton, $B_0$ is the external magnetic field, $k_B$ is Boltzmann's constant, $T$ is temperature, $\mu_0$ is the vacuum permeability, $g_n$ is the nuclear *g*-factor, $\mu_N$ is the nuclear magneton, $h$ is Planck's constant, and $\theta$ is the angle between the external magnetic field and the dipolar distance vector, $\mathbf{r}_{dip}$. King *et al.* verified Eq. (11) experimentally for a crystal of yttrium ethyl sulfate doped with Yb$^{3+}$ at temperatures below 3.8 K.[63]

Adding the bulk nuclear relaxation rate gives the following practical expression for the distance and angle dependence of the longitudinal relaxation rate of protons close to the trityl radical:

$$\frac{1}{T_{1n}} \simeq \left(\frac{\mu_0}{4\pi} g_e \mu_B \frac{1 - 3\cos^2\theta}{r_{dip}^3} \frac{1}{B_0}\right)^2 \frac{1}{T_{1e}} \text{sech}^2\left(\frac{g_e \mu_B B_0}{2 k_B T}\right) + \frac{1}{T_{1n,bulk}} \tag{13}$$

Figure 2 shows $T_{1n}$ as a function of the electron-nuclear distance for various angles $\theta$. The orientation dependence at 3.5 Å is plotted in Figure S1. Note that at Q band for $\theta = 0°$, $T_{1n}$ decreases from the bulk value of 52 s down to 0.19 s at 3.5 Å. However, at W band, $T_{1n}$ decreases only to 1.5 s.

The square around the orientation function in Eq. (13) gives it a fourth spherical rank and thereby puts it outside the common orientation-dependent line width formalisms, such as *g*-strain. The powder.m context function of *Spinach* kernel was therefore extended to allow the user to supply arbitrary angular dependences for the relaxation rates in the form of *Matlab* function handles.

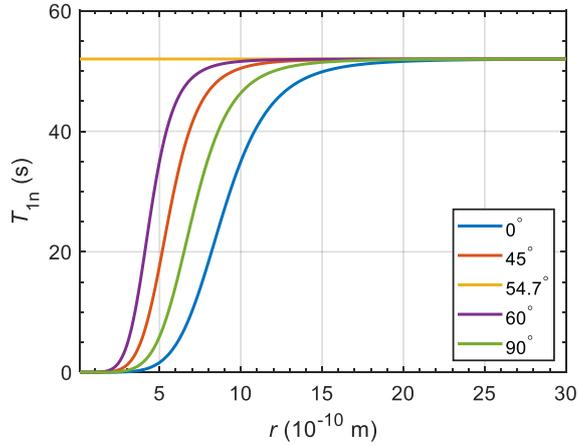

**Figure 2.** *Nuclear longitudinal relaxation time $T_{1n}$ as a function of electron-proton distance at Q band, following Eq. (13). The value of the angle $\theta$ between the direction of the external magnetic field and the dipolar distance vector is given in the legend. For these angles, the values of $T_{1n}$ are 0.19, 3.0, 52, 10, and 0.75 s, respectively, at 3.5 Å.*

### 3.5 Ensemble averaging

The exact location of nearby protons varies per trityl molecule in the frozen glassy matrix. As a consequence, a distribution of electron-proton dipolar couplings exists within the sample. To account for the effects of this distribution, we average the simulation outcome over electron-proton distances. Together with the usual system orientation averaging, this yields a triple integral:

$$\overline{\langle \mathbf{I}_Z \rangle_\infty} = \frac{3}{4\pi\left(R_{max}^3 - R_{min}^3\right)} \int_{R_{min}}^{R_{max}} r^2 dr \int_0^\pi \sin\theta d\theta \int_0^{2\pi} d\varphi \left\{ \langle \mathbf{I}_Z \rangle_\infty (r,\theta,\varphi) \right\} \quad (14)$$

with $R_{min} = 3.5$ Å as discussed above. $R_{max} = 20$ Å was empirically found to converge the radial part of the average. The angular part of this integral was evaluated using an 800-point REPULSION grid[64], and the radial part was done using the Gauss-Legendre quadrature[65]; both are implemented in *Spinach*.

On top of the orientation and distance distribution, we also consider a distribution in electron Rabi frequencies. Its physical origin is that the strength of the magnetic field component of the microwave irradiation ($|\mathbf{B}_1|$) is not uniform across the sample. The Gauss-Legendre quadrature[65] was used here, too. For the simulations of TOP and XiX DNP at Q band, we have averaged the simulation over a uniform distribution of electron Rabi frequencies between 10 and 20 MHz, unless noted otherwise.

## 4. Results

### 4.1 Performance benchmarks

As discussed in Section 2.2, the effective-Hamiltonian algorithm for computing the stroboscopic steady state is numerically unstable in dissipative DNP settings. Stable alternatives are propagator squaring and the Newton-Raphson method. We have benchmarked them; full results are presented in Figure S2 of the Supplementary Information. Propagator squaring and Newton-Raphson perform similarly up to the Liouville space dimension of 256 (*e.g.* one spin-1/2 electron and three protons without state-space restriction). For larger state spaces, Newton-Raphson is faster – up to three times for one electron and five protons. A GPU becomes advantageous also for matrix dimensions above 256: for a system with one electron and five protons with state space restriction up to four-spin correlations (propagator dimension 1909), our Nvidia A100 card was six times faster than multi-threaded matrix arithmetic on 32 Intel Xeon 6326 Gold CPU cores. GPU memory utilization should be carefully monitored – at the time of writing, sparse matrix arithmetic in CUDA libraries is unreasonably memory-hungry, this feature is unfortunately inherited by *Matlab*. Benchmarks are architecture-, hardware-, and implementation-dependent, but Newton-Raphson method on a GPU is generally recommended for stability and complexity scaling – it was used in all simulations discussed below.

**4.2 Field profiles**

Figure 3 shows simulations as well as the experimental field profile (top) of XiX DNP at Q band. Efficient DNP occurs at the microwave resonance offsets ±39 and ±60 MHz, less efficient DNP occurs at ±13 and ±81 MHz. The simulation of a single polarization transfer (bottom) predicts all these DNP conditions, but not with the correct intensities: most noticeable, the simulated peaks at ±13 MHz are too strong. The shapes of the peaks are not correctly predicted either: in the single-transfer simulation, the peaks are pointy and asymmetric, reflecting the axial *g*-anisotropy of trityl, whereas in the experiment they are smooth and more symmetric. In the experiment, some enhancement is observed in between the matching conditions, for example, at resonance offsets of ±50 MHz, but this is not reproduced in the single-transfer simulation. The steady-state simulations show improvements on all these fronts, even when just a single parameter set is considered. Further improvement is obtained when distributions of distances and $B_1$-fields are integrated over. When the ensemble of electron Rabi frequencies is included, the weak intensities of the DNP conditions at ±13 MHz are correctly predicted. These conditions are active for a narrow range of electron Rabi frequencies (18-20 MHz),[22] which makes them susceptible to the – apparently – quite inhomogeneous microwave $B_1$-field across the experimental sample. The other matching conditions are active for a broader range of Rabi frequencies, they therefore remain unaffected by the $B_1$ inhomogeneity.

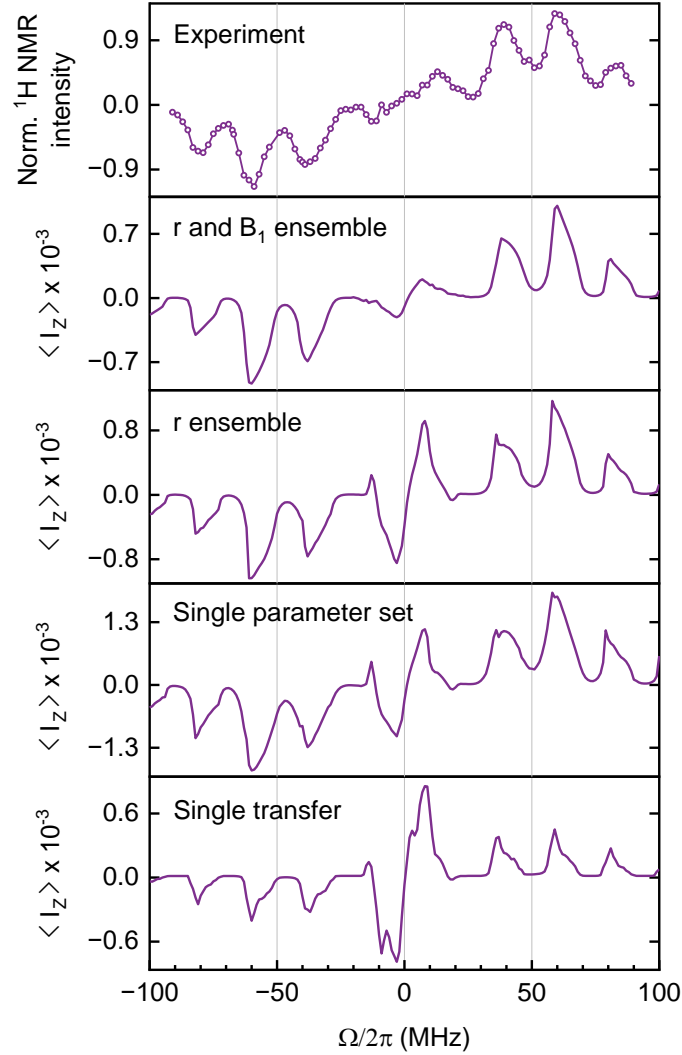

*Figure 3.* Field profiles of XiX DNP at Q band. From bottom to top: simulation of a single polarization transfer; steady-state simulation with a single set of parameters; steady-state simulation with the electron-proton distance ensemble; steady-state simulation with the electron-proton distance and the electron Rabi frequency ensembles; experiment. The parameters of the XiX DNP experiment were: $V_{1S}$ = 18 MHz, $t_p$ = 48 ns, $t_c$ = 3456 ns (36 blocks), $t_{rep}$ = 204 µs, and $^1$H NMR recycle delay = 24 s.

Figure 4 shows simulations as well as the experimental field profile (top) of XiX DNP at W band. Enhancements at the microwave resonance offsets at ±110 and ±170 MHz dominate. The single-transfer simulation predicts this fairly well, but the shapes of the peaks are again not correct. Some improvement is accomplished with the steady-state simulation based on a single parameter set, but the shapes are still off and intensities at resonance offsets ±60 and ±230 MHz come out too strong. Both problems are corrected when the electron-proton distance ensemble is included; the largest distance in the ensemble (20 Å) was adjusted to reproduce the relative peak intensities in the experimental field profile. A strong dependence on the electron Rabi frequency is not expected for any of the active DNP conditions (Figure S3). Therefore, including a $B_1$ ensemble has no significant effect.

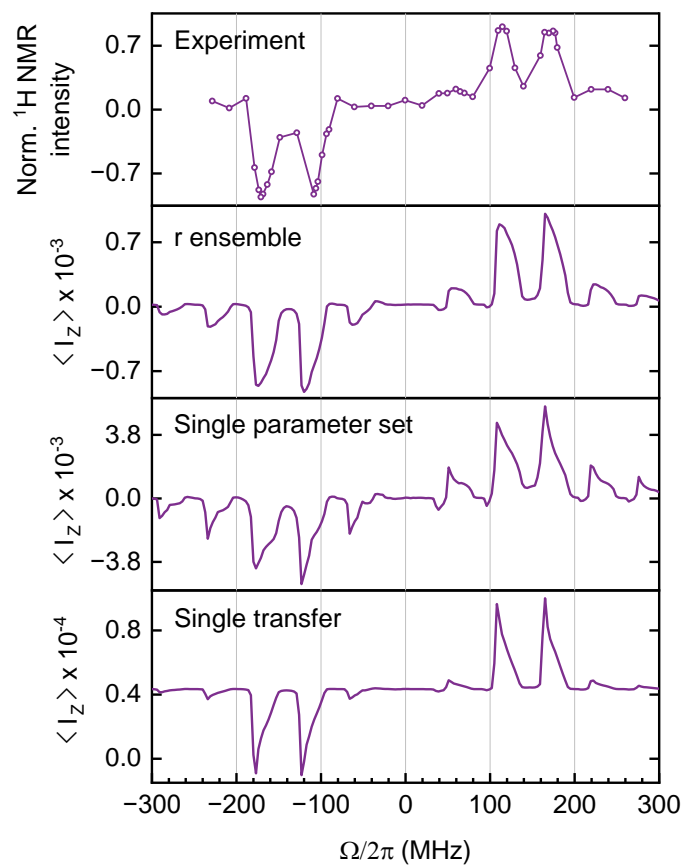

*Figure 4.* Field profiles of XiX DNP at W band. From bottom to top: simulation of a single polarization transfer; steady-state simulation with a single set of parameters; steady-state simulation with the electron-proton distance ensemble; experiment. Before each $^1$H NMR acquisition in the experimental field profile, the XiX DNP sequence was applied for 60 s, with $v_{1S}$ = 20 MHz, $t_p$ = 18 ns, $t_c$ = 360 ns (10 blocks), and $t_{rep}$ = 167 µs.

Parameter scans are a prerequisite for finding the best conditions for the TOP/XiX/TPPM/BEAM DNP sequences. Previously[22] we relied on single-transfer simulations for this purpose (see Figure S3 for an example at W band). In Figure 5, a steady-state simulation and experimental parameter scan are compared, for the purpose of optimizing the pulse length in XiX DNP at W band. Just as with the field profiles, the agreement with experiment is much improved, especially regarding the shapes and relative intensities of the DNP conditions. The steady-state simulation took just over an hour to compute.

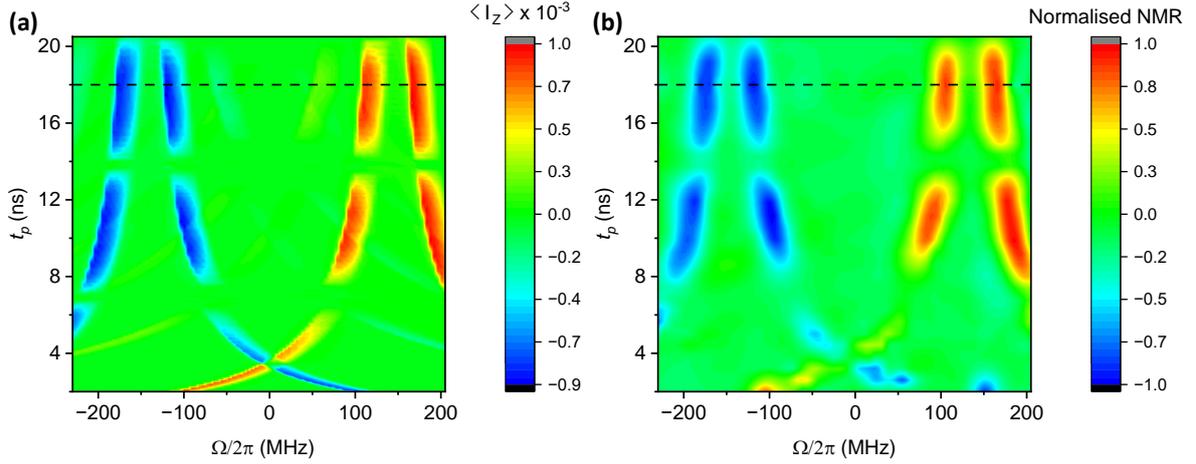

*Figure 5. Parameter scans for XiX DNP at W band: **(a)** steady-state simulation with the electron-proton distance ensemble and **(b)** experiment. To obtain the experimental parameter scan, the $^1$H NMR signal was acquired for a total of 460 combinations of microwave resonance offset and pulse length. Sampling was denser in regions where strong enhancement was expected and data points were interpolated onto a uniform parameter grid for plotting. The XiX DNP sequence was applied for 60 s, with $\nu_{1S}$ = 20 MHz and $t_{rep}$ = 167 µs. Contact times were adjusted between 340 and 370 ns to accommodate complete XiX blocks. To obtain the simulated parameter scan, 20200 combinations of microwave resonance offset and pulse length were computed. The dashed horizontal lines in both plots indicate the corresponding field profiles in Figure 4.*

### 4.3 Repetition-time profiles

The repetition time at which the highest enhancement factor is achieved, strongly depends on the choice of the DNP pulse sequence.[22] Simulation of a single transfer of polarization cannot predict this property, but a steady-state simulation can, as shown in Figure 6. As with the field profiles, including electron-proton distance and electron Rabi frequency ensembles further improves the agreement with experiments (Figure S4).

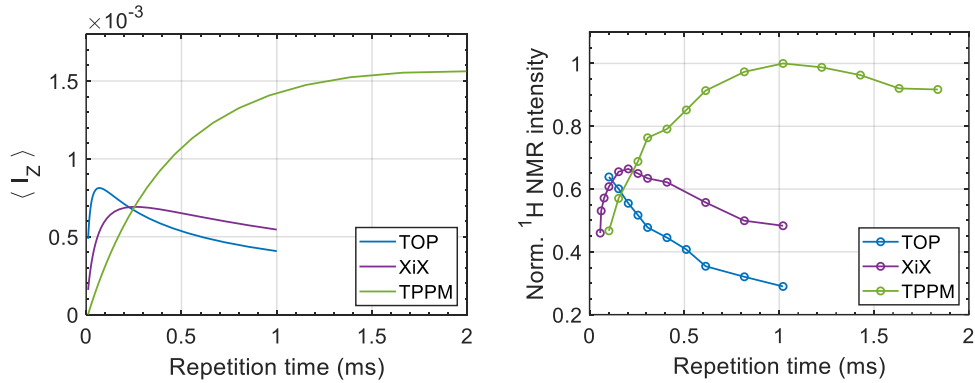

*Figure 6. Optimization of repetition times for TOP, XiX, and TPPM DNP at Q band. Left: steady-state simulations with electron-proton distance and electron Rabi frequency ensembles, right: experiments. The parameters for the TOP DNP experiment were: $\nu_{1S}$ = 18 MHz, $t_p$ = 10 ns, $d$ = 14 ns, $t_c$ = 7200 ns, and $\Omega/2\pi$ = 95 MHz; for XiX: $\nu_{1S}$ = 18 MHz, $t_p$ = 48 ns, $t_c$ = 3456 ns, and $\Omega/2\pi$ = –39 MHz; for TPPM: $\nu_{1S}$ = 33 MHz, $t_p$ = 16 ns, $\phi$ = 115°, $t_c$ = 9600 ns, and $\Omega/2\pi$ = 2 MHz. In the simulations, the range of the electron Rabi frequencies is 10-20 MHz for TOP and XiX DNP and 25-35 MHz for TPPM DNP. For XiX, the absolute values of $\langle \mathbf{I}_Z \rangle$ and the $^1$H NMR intensity are plotted for clarity.*

The same set of relaxation parameters (with the exception of $T_{1n,bulk}$, which was adjusted to 26 s following the experiments) also reproduces the repetition-time profile of NOVEL at X band, see Figure 7. When a flip-back pulse is applied after the contact time,[19,51] the optimal repetition time shortens from about 1 ms to 0.5 ms and the peak enhancement increases. Interestingly, in the steady-state simulation of NOVEL with flip-back, the predicted nuclear polarization peaks above the maximum theoretical enhancement factor of 329.

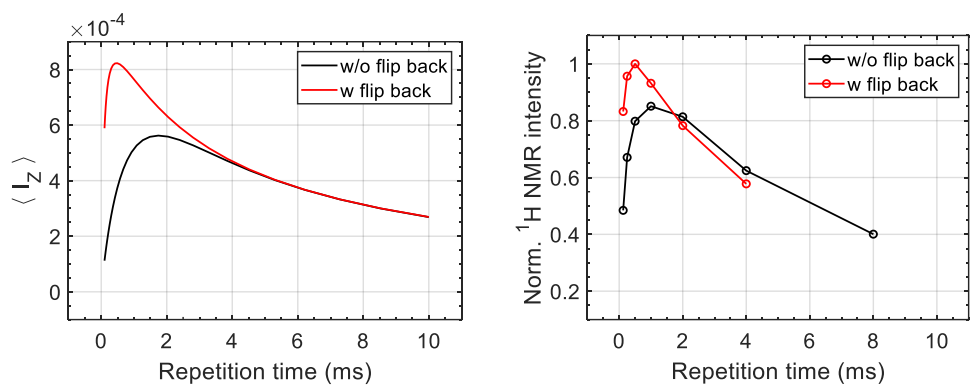

*Figure 7. Optimization of repetition times for NOVEL, with and without the flip-back pulse after the contact time, at X band. Left: steady-state simulations with the electron-proton distance and the electron Rabi frequency ensembles, right: experiments with the following parameters: $v_{1S}$ = 15 MHz, $t_c$ = 500 ns, and $\Omega/2\pi$ = 0 MHz. In the simulations, the range of the electron Rabi frequencies is 14-16 MHz.*

### 4.4 Polarization transfer during the DNP sequence

Figure 8 shows the polarization transfer as a function of the contact time, $t_c$. The single transfer simulations in Figure 8a do not predict the experiments in Figure 8c very well. Prominent oscillations remain (even after powder averaging), because the dipolar coupling to only two proton positions is considered, while the experiments show a smooth transfer curve. Worse is that the transfer times are incorrectly predicted for both XiX (purple) and high-power TOP DNP (red): in the experiment, transfers with these sequences are about as fast as with TPPM, but the simulations predict them to be much slower. Once again, steady-state simulations are more accurate (Figure 8b). Inclusion of the electron-proton distance ensemble smoothens the oscillations and inclusion of the electron Rabi frequency ensemble slows down the transfer under low-power TOP (blue) (Figure S5), resembling experiments. Prediction of relative enhancement factors also improves, but is not yet perfect, at least in part because the $^1$H NMR signals were acquired with a recycle delay of 24 s, which is well below five times the bulk build-up times for all sequences.[22] Finally, the steady-state simulations show a gradual decay of the nuclear polarization for contact times longer than 1 μs, while the experiments show a gradual increase of the $^1$H NMR signal on this time scale for all sequences except XiX. A likely explanation is that our modelling is still not complete; we come back to this point in the Discussion. Note that the crossovers between XiX (purple) and TOP (red and blue) are reproduced.

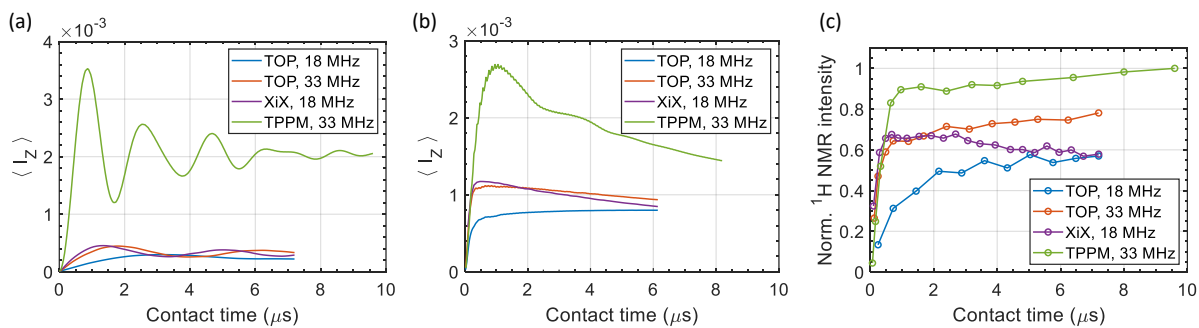

*Figure 8. Transfer of longitudinal magnetization during the TOP, XiX, and TPPM DNP pulse sequences at Q band. **(a)** Single-transfer simulations, from one electron to two protons, **(b)** steady-state simulations with the electron-proton distance and the electron Rabi frequency ensembles, and **(c)** experiments. The parameters for the TOP DNP experiments were: $\nu_{1S}$ = 18 or 33 MHz, $t_p$ = 10 ns, $d$ = 14 ns, $t_{rep}$ = 102 or 153 µs, and $\Omega/2\pi$ = 95 or 92 MHz; for XiX: $\nu_{1S}$ = 18 MHz, $t_p$ = 48 ns, $t_{rep}$ = 153 µs, and $\Omega/2\pi$ = 61 MHz; for TPPM: $\nu_{1S}$ = 33 MHz, $t_p$ = 16 ns, $\phi$ = 115°, $t_{rep}$ = 816 µs, and $\Omega/2\pi$ = 2 MHz.*

### 4.5 Electron Rabi frequency dependence

Redrouthu *et al.* observed that the optimal repetition time not only depends on the DNP pulse sequence, but also on the electron Rabi frequency.[22] This implies that single-transfer simulations cannot provide reliable information about the dependence of the enhancement factor on the electron Rabi frequency. Figure 9 shows that this problem is also tackled by the steady-state simulations. In the experiments, the repetition time and the resonance offset were optimized for each electron Rabi frequency. The result was an almost constant enhancement factor with XiX DNP, in contrast to a steadily increasing enhancement factor with TOP DNP. This remarkable behavior is reproduced in the steady-state simulations, including the gradual shift of the resonance offsets.

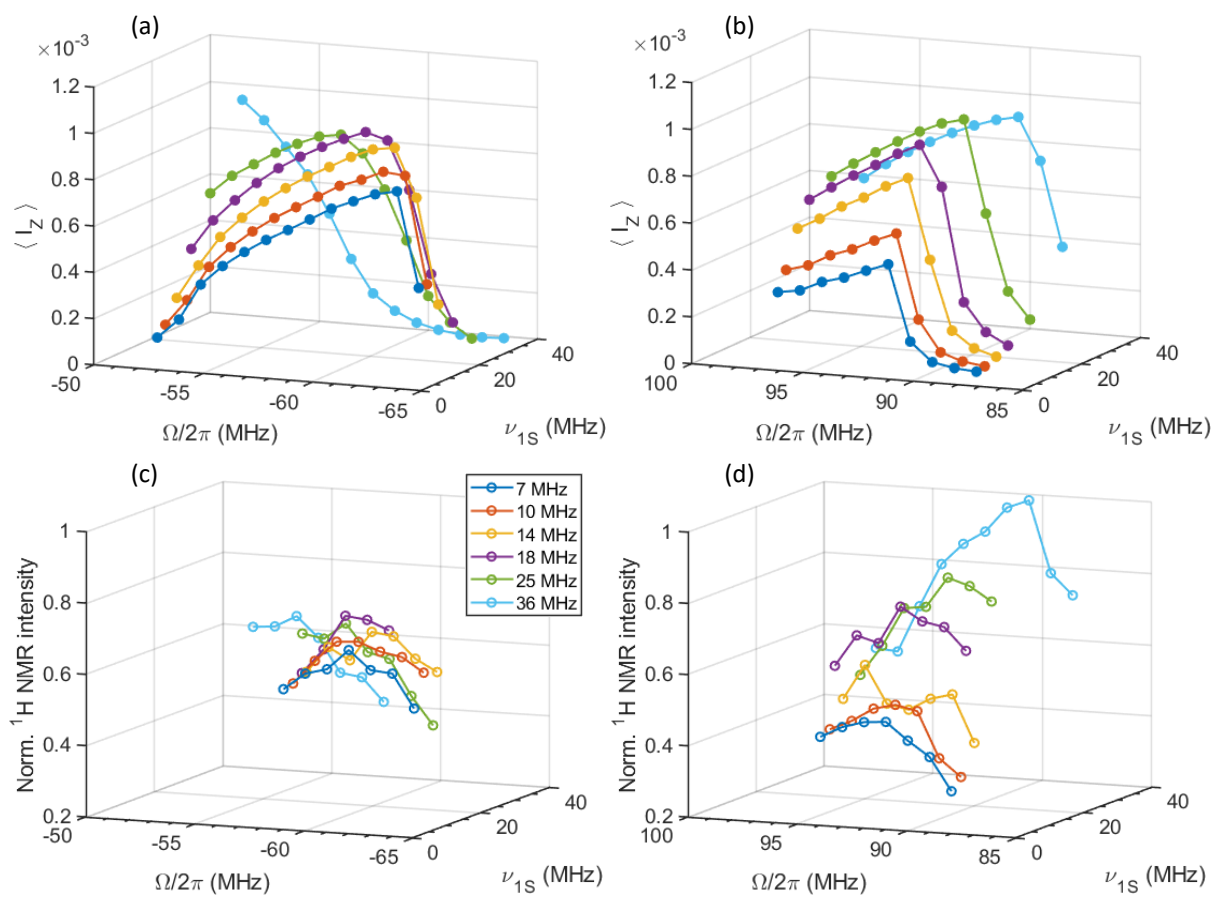

*Figure 9. Dependence of the enhanced nuclear polarization on the electron Rabi frequency for XiX and TOP DNP at Q band. (a,b) Steady-state simulations with the electron-proton distance and the electron Rabi frequency ensembles. For $\nu_{1S}$ = 7 MHz, the ensemble range is 1-9 MHz. For the other values of $\nu_{1S}$, the range is $\nu_{1S}$ –8 to $\nu_{1S}$ +2 MHz. (c,d) Corresponding experiments for XiX and TOP DNP. For XiX, the optimal repetition time increases from 51 to 306 μs as the electron Rabi frequency increases, whereas for TOP, the optimal repetition time remains constant at 153 μs. For a full account of all parameters, see the Supplementary Materials of Redrouthu et al.[22]*

## 5. Discussion

The steady-state simulation methods introduced above have enabled accurate prediction of pulsed-DNP field profiles, optimal repetition times, polarization transfer during the sequence, and the dependence on the electron Rabi frequency. Remarkably, it was sufficient to consider just one electron and one proton together with distributions in the electron-proton distances and microwave $B_1$-fields. For the nuclear relaxation, we used a crude existing model.[61] This worked: the different pulsed-DNP experiments were successfully simulated with the same set of parameters. While improvements are certainly possible, as we discuss below, simulation of pulsed DNP in the steady state is a clear step forward compared to the single-transfer simulations.

To make sure that the relaxation parameters are not wildly off, we investigated the effects of alternative values of $T_{1e}$, $T_{2e}$, $T_{1n}$, $T_{2n}$ on the field and repetition-time profiles and the polarization transfer during the sequence for XiX DNP at Q band (Figure S6). As expected, increasing $T_{2e}$ and $T_{1n}$ leads to higher enhancement factors, while increasing $T_{1e}$ leads to lower enhancement factors. Changes in $T_{2n}$ have only moderate effects. Given the simplicity of the spin system, we do not expect that the current simulations can quantitatively predict enhancement factors. This makes that the shapes of the repetition-time profiles, more so than the field profiles, are helpful to validate the relaxation modelling. The simulations in black (Figure S6) were obtained with the parameters as described in the Materials and Methods section; we have found at least a local minimum. Further investigation is, however, warranted, particularly concerning $T_{2n}$. Effects of including the distance and orientation dependence of $T_{1n}$ are barely noticeable in the simulations in Figure S6 (overlapping black and red curves). The reason is that for a large fraction of the protons in the electron-proton distance distribution, $T_{1n}$ remains at the bulk value. In simulations with one electron and one proton at 3.5 Å, including the distance and orientation dependence of $T_{1n}$ decreases the enhancement by a few percent.

The obvious next step is to include more nuclei in the spin system. We expect that this will fix the remaining shortcomings in the simulations of the transfer during the DNP sequence (Figure 8), which seem to relate to a fraction of protons that polarize more slowly than the rest. Modelling of proton-proton spin diffusion will allow us to simulate the build-up times of the bulk polarization and bring us closer to quantitative prediction of the enhancement factors. Large-scale *ab initio* simulations have already been demonstrated for the cross-effect under MAS[35] and we can follow the same approach. Briefly, this means that we need a realistic set of nuclear locations in 3D and use the neighbor's cutoff[34] to prune the restricted basis set further. With this in place, theoretical optimization becomes possible for both the DNP pulse sequence and the chemical structure of the polarizing agent.[66] Going further, effects of electron-electron interactions should be modelled as well.[67] Simulation of pulsed DNP under MAS will produce complications of its own, but software infrastructure to handle this exists.[43,68]

The rate of the polarization transfer is determined by the electron-proton dipolar coupling and its preservation in the effective Hamiltonian.[22] This is a property of the DNP pulse sequence and varies per DNP condition (choice of the pulse length, resonance offset, electron Rabi frequency, etc.). Generally, a higher electron Rabi frequency means a better preservation, as was recently analytically shown.[69] A faster rate of transfer leads to a faster build-up of the bulk polarization, but not necessarily to a larger enhancement; other factors, such as the response of the pulse sequence to the width of the EPR spectrum of the polarizing agent, play a role as well. To derive the effective Hamiltonian, perturbation theory is used to capture the trajectory of the spin system in a suitable interaction frame. Analytical treatments of this kind are in perfect agreement with numerical simulations of the single transfers.[22] However, the reality of a pulsed DNP experiment, which always requires many transfers, is more complex. For XiX and TOP DNP, the polarization is transferred considerably faster than analytical theory predicts (Figures 8a,c) and, interestingly, the steady-state simulations reproduce this very well (Figure 8b). This implies that the steady orbit differs from the trajectory during the single (one-time) transfer.

Small wiggles are observed in the steady-state simulations of the polarization transfer (Figure 8b), but not in the single-transfer simulations (Figure 8a). They smoothen with ensemble averaging (Figure S5). We have excluded numerical noise as the cause. Instead, they arise because the direction of the electron spin magnetization vector at the end of the DNP pulse sequence (slightly) varies with the number of blocks. As a consequence, after a period of relaxation that is clearly shorter than five times $T_{1e}$, the starting direction before the renewed application of the DNP pulse sequence changes, differently depending on the number of blocks. Thus, the wiggles reflect that the steady orbit depends on the contact time (and the repetition time). This is particularly the case for XiX and high-power TOP DNP with short contact times, as becomes clear from the unexpectedly large values of $\langle \mathbf{I}_Z \rangle$. These steady orbits enable a better preservation of the electron-proton dipolar coupling than expected from analysis of the sequence and the electron Rabi frequency. Optimization of the effective Hamiltonian has recently been used to develop the broadband DNP pulse sequence PoLarizAtion Transfer via non-linear Optimization (PLATO).[70] The effective Hamiltonian of a single transfer was calculated numerically using the matrix logarithm, in Hilbert space. Extension of this optimization approach to the steady state, however, requires further thought, for starters because of the numerical instability of the matrix logarithm method in Liouville space. Implementations of gradient ascent pulse engineering (GRAPE)[71] for dissipative evolution generator design also do not yet exist. Finding a way to engineer these fast steady orbits is nevertheless important, because they might provide a loophole to efficient high-field pulsed DNP at low peak power.

The ensemble averaging in this work was over the electron-proton distance and the electron Rabi frequency. These are not the only distributed quantities in a glassy DNP matrix. The flexibility of the chemical linker in polarizing agents for cross-effect DNP is a well-known source of distributions in the electron-electron dipolar couplings, the exchange interaction, and the relative positions and orientations of the radical moieties.[72] There are also distributions in the distance to the next nearest polarizing agent, internuclear distances depend on the conformations, etc. If *d*- or *f*- elements are used as polarizing agents, magnetic properties vary drastically with conformation,[73] which would in turn affect the relaxation of nearby nuclei. All together there are far too many parameters for a product grid integral to be affordable; Monte-Carlo averaging would have to be used instead.

## 6. Conclusion

We have introduced the simulation of pulsed DNP in the steady state. Several algorithms to compute the stroboscopic steady state from the pulsed-DNP propagator have been explored; of these, we recommend the Newton-Raphson method for its stability and complexity scaling. Agreement with experiments at X, Q, and W band, with DNP pulse sequences NOVEL, TOP, XiX, and TPPM, is remarkably good, even when a spin system of just one electron and one proton is considered. The steady-state simulations are expected to be equally useful for DNP pulse sequences that induce polarization transfer by adiabatic passage. An important observation is that the trajectory of the spin system during the first application of the DNP pulse sequence differs from the steady orbit. This may provide a clue to designing efficient pulse sequences for high-field MAS DNP. We will put the steady-state simulations to use exactly for this purpose.

## Acknowledgement

This project was funded by the Deutsche Forschungsgemeinschaft through SFB 1527 (project-ID 454252029) and a research grant from the Center for New Scientists at the Weizmann Institute of Science. The authors thank Robert Hunter and Hassane El Mkami for assistance with the W-band experiments.

## Data and code availability

All data generated in this study has been deposited in the Zenodo repository. Scripts for numerical simulations are available in the example set of the *Spinach* library.

SUPPLEMENTARY INFORMATION

# Simulation of pulsed dynamic nuclear polarization in the steady state

Shebha Anandhi Jegadeesan[1], Yujie Zhao[2,3], Graham M. Smith[2], Ilya Kuprov[4,5], and Guinevere Mathies[1,*]

[1]Department of Chemistry, University of Konstanz, Konstanz, Germany

[2]School of Physics and Astronomy, University of St Andrews, St Andrews, United Kingdom

[3]Current address: Francis Bitter Magnet Laboratory and Department of Chemistry, Massachusetts Institute of Technology, Cambridge, Massachusetts, United States

[4]Chemical and Biological Physics Department, Weizmann Institute of Science, Rehovot, Israel

[5]School of Chemistry and Chemical Engineering, University of Southampton, Southampton, United Kingdom

## Contents

*Email: guinevere.mathies@uni-konstanz.de

# 1. Orientation dependence of $T_{1n}$ near an unpaired electron

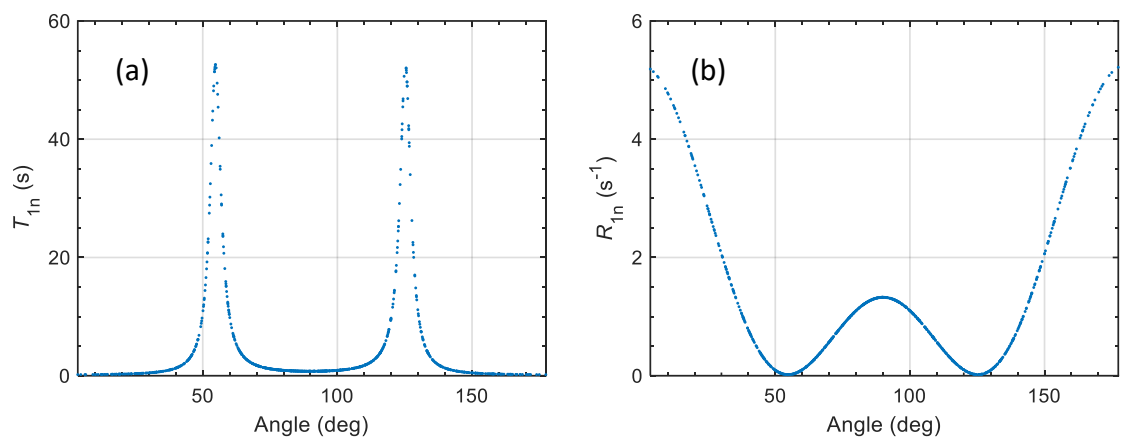

*Figure S1.* Longitudinal relaxation *(a)* times and *(b)* rates for a spin-1/2 nucleus at 3.5 Å from an unpaired electron, plotted as a function of the angle between the dipolar vector and the external magnetic field (Eq. (13) from the main manuscript). Close to 0° and 180°, $T_{1n}$ is 191 ms; close to 90°, $T_{1n}$ is 754 ms. At the magic angle, 54.7° (and 125.3°), $T_{1n}$ equals the bulk value of 52 s.

# 2. Performance benchmarks

Benchmarking of the steady-state algorithms was conducted on a Dell T550 server equipped with 32 Intel Xeon 6326 CPU cores, 512 GB of RAM, and an Nvidia A100 GPU (80 GB of RAM, 9.7 TFLOPS FP64). As a representative case, two points of the XiX DNP field profile at Q band were computed for 100 points in the spherical orientation grid ($t_c$ = 3072 ns, $t_{rep}$ = 204 μs), including orientation and distance dependence of $T_{1n}$.

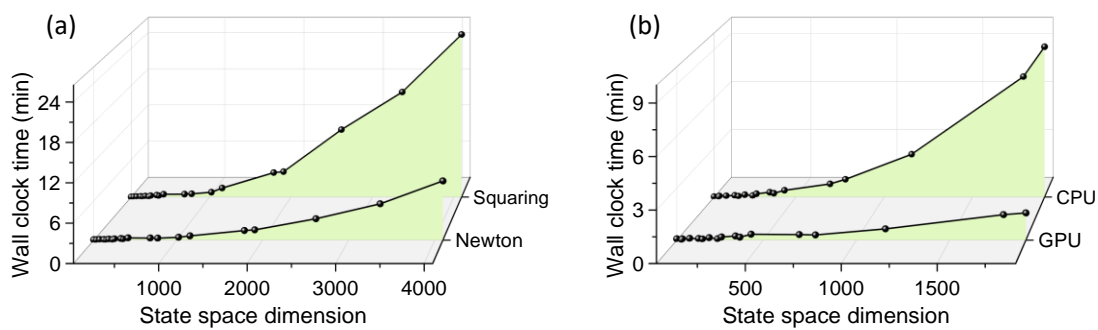

*Figure S2.* Wall clock calculation time (see Table S1 for numerical values) as a function of state space dimension for (a) propagator squaring vs. Newton-Raphson (both using the GPU) and (b) Newton-Raphson for GPU vs. CPU.

*Table S1.* *Raw timing data for the plots in Figure S2.*

| Number of protons | Max. spin correlation level | State space dimension | Wall clock time (s) | | |
|---|---|---|---|---|---|
| | | | Squaring GPU | Newton GPU | Newton CPU |
| 2 | 2 | 37 | 2 | 1 | 2 |
| 2 | 3 | 64 | 1 | 1 | 1 |
| 3 | 2 | 67 | 3 | 4 | 3 |
| 3 | 3 | 175 | 3 | 3 | 3 |
| 3 | 4 | 256 | 5 | 5 | 5 |
| 4 | 2 | 106 | 4 | 6 | 4 |
| 4 | 3 | 376 | 13 | 10 | 14 |
| 4 | 4 | 781 | 31 | 18 | 65 |
| 4 | 5 | 1024 | 47 | 26 | - |
| 5 | 2 | 154 | 6 | 6 | 6 |
| 5 | 3 | 694 | 28 | 19 | 48 |
| 5 | 4 | 1909 | 246 | 94 | 550 |
| 5 | 5 | 3367 | 1018 | 333 | - |
| 5 | 6 | 4096 | 1576 | 543 | - |
| 6 | 2 | 211 | 9 | 9 | 9 |
| 6 | 3 | 1156 | 87 | 39 | 157 |
| 7 | 2 | 277 | 13 | 11 | 12 |
| 7 | 3 | 1789 | 237 | 88 | 440 |
| 8 | 2 | 352 | 18 | 15 | 17 |
| 8 | 3 | 2620 | 653 | 196 | - |
| 9 | 2 | 436 | 27 | 20 | 25 |

## 3. Dependence on electron Rabi frequency at W band

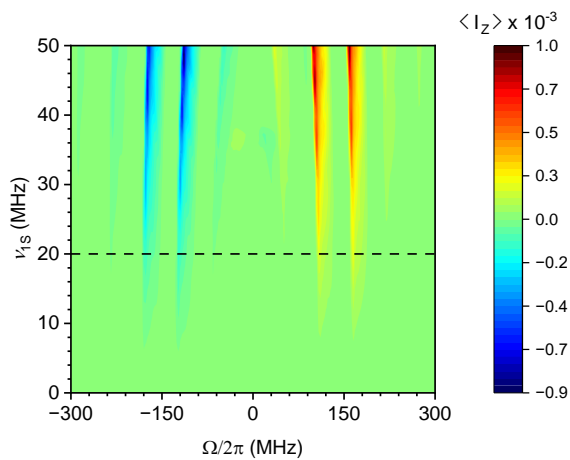

**Figure S3.** Single-transfer parameter scan for XiX DNP at W band with $t_p$ = 18 ns and $t_c$ = 1000 ns. The spin system consists of one unpaired electron at the origin and two protons at (0.0, 3.5, 0.0) and (2.475, 2.475, 0.0) Å. g-anisotropy is included.

## 4. Repetition-time profiles

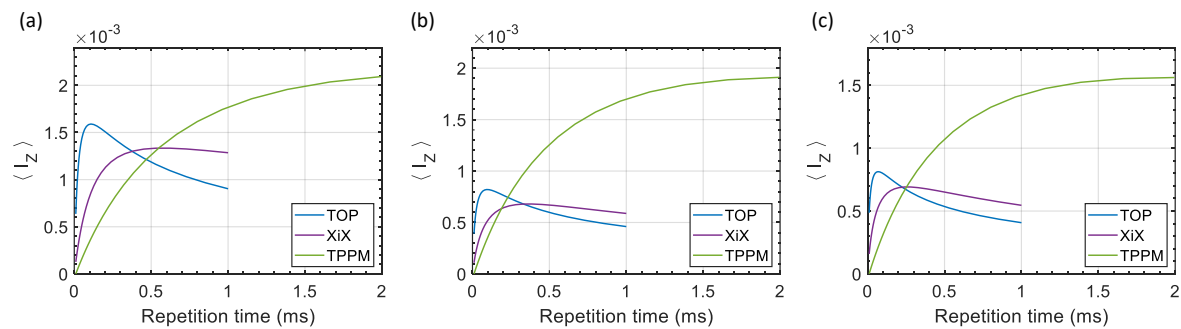

**Figure S4.** Optimization of repetition times for TOP, XiX, and TPPM DNP at Q band. Steady-state simulations with **(a)** a single parameter set, **(b)** electron-proton distance ensemble, and **(c)** electron-proton distance and electron-Rabi-frequency ensembles. The electron-proton distance ensemble has a particularly detrimental effect on the XiX and TOP DNP enhancement factors and thereby brings the relative enhancement factors in better agreement with the experiments (Figure 6 in the main manuscript).

## 5. Polarization transfer during the DNP sequence

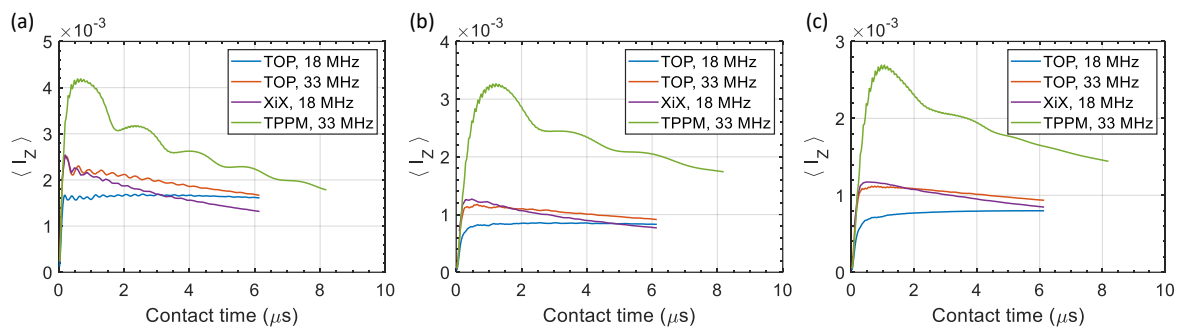

*Figure S5.* Transfer of longitudinal magnetization during the TOP, XiX, and TPPM DNP pulse sequences at Q band. Steady-state simulations with **(a)** a single parameter set, **(b)** electron-proton distance ensemble, and **(c)** electron-proton distance and electron Rabi frequency ensembles. In the simulations with a single parameter set **(a)**, the oscillations during the initial rise are still present, but they are smoothed out by the ensemble averaging **(b,c)**.

## 6. Effects of alternative relaxation times

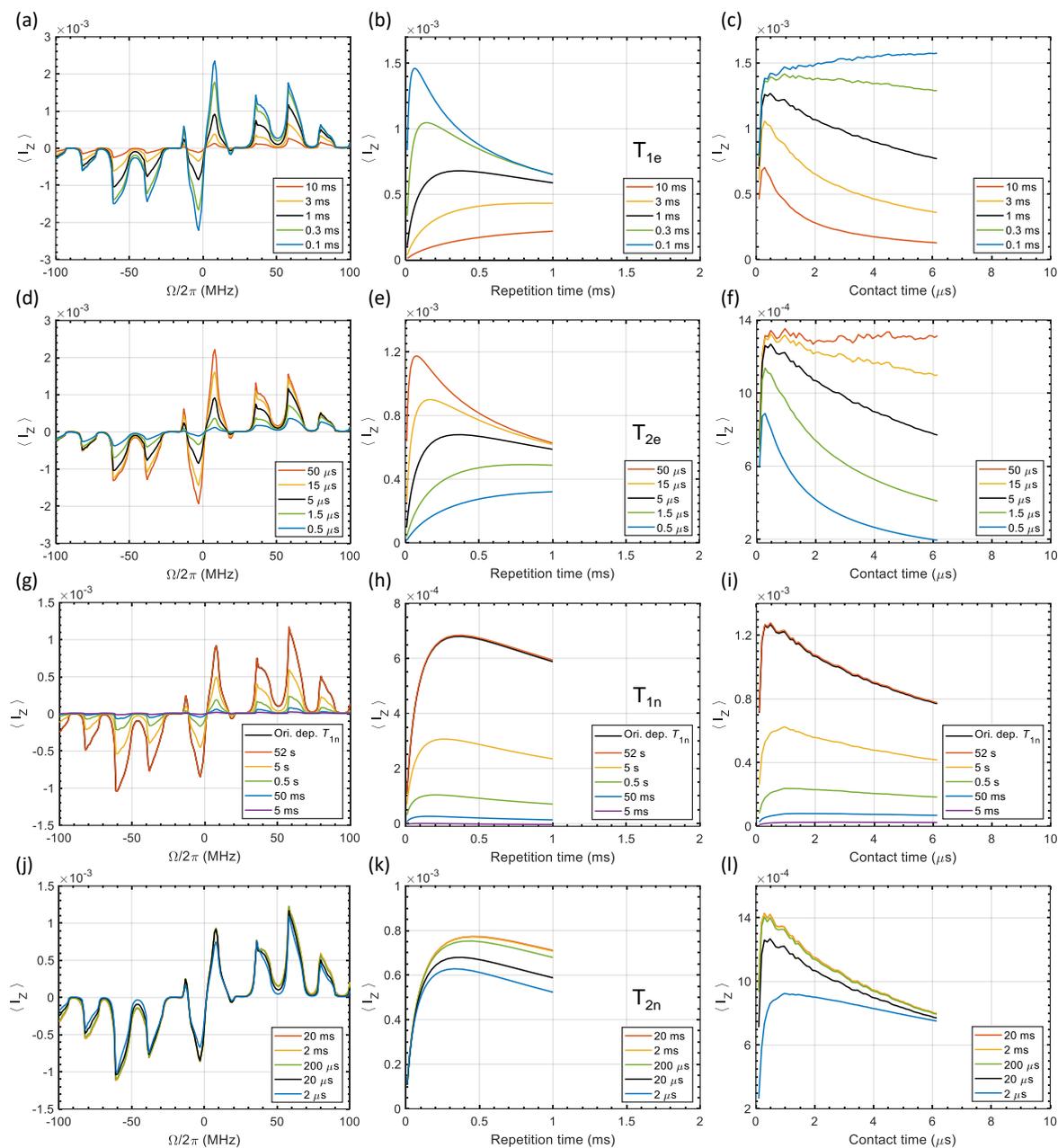

*Figure S6.* Effects of alternative electron and nuclear spin relaxation times on field profiles, repetition-time profiles, and polarization transfer for XiX DNP at Q band. **(a-c)** Electron spin longitudinal relaxation, **(d-f)** electron spin transverse relaxation, **(g-i)** nuclear spin longitudinal relaxation, varied without distance and orientation dependence, and **(j-l)** nuclear spin transverse relaxation. Black curves are simulations with the relaxation parameters of the main manuscript. All simulations include the electron-proton distance ensemble.